\documentclass[12pt]{JHEP3}

\usepackage{amsmath}
\usepackage{amssymb}
\usepackage{graphicx}

\newcommand{\be}{\begin{eqnarray}}
\newcommand{\ee}{\end{eqnarray}}

\newcommand{\bn}{\begin{enumerate}}
\newcommand{\en}{\end{enumerate}}

\def\Tr{{\rm Tr}}

\title{Index computation for 3d Chern-Simons matter theory: test of
Seiberg-like duality }

\author{ Chiung Hwang $^{1}$, Hyungchul Kim$^{1}$ , Kyung-Jae Park$^{1}$, Jaemo Park$^{1,2}$

\\

\\

$^1$Department of Physics, POSTECH, Pohang 790-784, Korea
\\
$^2$Postech Center for Theoretical Physics (PCTP), Postech, Pohang
  790-784, Korea

\\
\\
E-mail: \email{ilvhemos@postech.ac.kr, dakiro@postech.ac.kr,
jaco@postech.ac.kr, jaemo@postech.ac.kr} } 

\abstract{ We work out the superconformal index for $\mathcal N=2$
supersymmetric Chern-Simons matter theories exhibiting Seiberg-like
dualities proposed by Giveon and Kutasov. We consider
$U(N)/Sp(2N)/O(N)$ gauge theories of QCD type and find the perfect
agreements for proposed dual pairs. }

\begin{document}


\section{Introduction}

Recently there have been tremendous progress in understanding of
three-dimensional superconformal field theories (SCFT). The key
observation was made by J. Schwarz that such theories could be
described as Chern-Simons matter theories \cite{Schwarz04}. This led
to the important development in ${\rm AdS}_4$/${\rm CFT}_3$
correspondence for the supersymmetric theories with $\mathcal N\geq
4$ \cite{BL1, BL2, BL3, gus1, gus2, GaiottoWitten,
Hosomichi08,Aharony08, Hosomichi08a, Imamura08}. However the same
insight can be used to understand the SCFT with $\mathcal N=2$
supersymmetry \cite{Gaiotto07}. For these theories, there have been
intense studies in the context of ${\rm AdS}_4$/${\rm CFT}_3$
correspondence \cite{Hanany1, Hanany2, Hanany3, Martelli,
Jafferis09, Benini10}. We are interested in a subset of such
theories, i.e., three-dimensional supersymmetric QCD with
Chern-Simons couplings. In the IR limit, the Yang-Mills kinetic term
is irrelevant and we are left with $\mathcal N=2$ Chern-Simons
matter theories. $\mathcal N=1$ supersymmetric QCD in
four-dimensions was intensively studied in relation to Seiberg
duality \cite{Seiberg95}. Down to three dimensions there's an
analogue of the Seiberg dualities in Chern-Simons matter theories
with $\mathcal N=3, \mathcal N=2$ supersymmetry \cite{Giveon09,
Niarchos}. Some of the evidences were presented in
\cite{Kapustin11,Willett11}, evaluating the partition function on
$S^3$. The purpose of the paper is to give additional evidences by
working out the superconformal index for dual pairs with $\mathcal
N=2$ supersymmetry. The index computation gives detailed information
of BPS states of the SCFT of interest. Indeed the index matches
perfectly and this provides a strong evidence that Seiberg-like
duality holds for three-dimensional $\mathcal N=2$ super
Chern-Simons matter theories of QCD type. The superconformal index
for QCD type theory without Chern-Simons term is computed by
\cite{Bashkirov}.

The content of the paper is as follows. After introducing the
essentials of superconformal index in three-dimensions, we appply
this for $\mathcal N=2 \,\, U(N), Sp(2N), O(N)$ Chern-Simons  theories with
fundamental matters. It's important to have the gauge group $U(N),
O(N)$ instead of $SU(N), SO(N)$ to have valid Seiberg-like
dualities. In the main text, we just keep track of the energy of the
state while in the appendix we turn on the chemical potentials for
the flavor symmetries and redo the index computation.

\section{Computation of the superconformal index}

Let us discuss the general structures of the index. We consider the
superconformal index for 3-d $\mathcal{N}=2$ superconformal field
theory (SCFT). Superconformal index for higher supersymmetric theory
can be defined using their $\mathcal{N}=2$ subalgebra. The bosonic
subgroup of the 3-d $\mathcal{N}=2$ superconformal algebra is
$SO(2,3) \times SO(2) $.  There are three Cartan elements denoted by
$\epsilon, j_3$ and $R$ which come from three factors
$SO(2)_\epsilon \times SO(3)_{j_3}\times SO(2)_R $ in the bosonoic
subalgebra.  One can define the superconformal index for 3-d
$\mathcal{N}=2$ SCFT as follows \cite{Bhattacharya09},
\begin{equation}
I=\Tr (-1)^F \exp (-\beta'\{Q, S\}) x^{\epsilon+j_3}\prod_j
y_j^{F_j} \label{def:index}
\end{equation}
where $Q$ is a special  supercharge with quantum numbers $\epsilon =
\frac{1}2, j_3 = -\frac{1}{2}$ and $R=1$ and $S= Q^\dagger$. They
satisfy following anti-commutation relation,
\begin{equation}
 \{Q, S\}=\epsilon-R-j_3 : = \Delta.
\end{equation}
In the index formula, the trace is taken over gauge-invariant local
operators in the SCFT defined on $\mathbb{R}^{1,2}$ or over states
in the SCFT on $\mathbb{R}\times S^2$. As is usual for Witten index
, only BPS states satisfying the bound $\Delta =0 $ contributes to
the index and the index is independent of $\beta'$. If we have
additional conserved charges commuting with chosen supercharges
($Q,S$), we can turn on the associated chemical potentials and the
index counts the number of BPS states with the specified quantum
number of the conserved charges denoted by $F_j$ in eq.
(\ref{def:index}).

The superconformal index is exactly caculable using localization
technique \cite{Kim09,Imamura11}.  Following their works, the
superconformal index can be written in the following form,
\begin{equation}\label{index}
I(x)=\sum_{m} \int da\, \frac{1}{(symmetry)}
e^{S^{(0)}_{CS}}e^{ib_0(a)} y_{j}^{q_{0j}}
x^{\epsilon_0}\exp\left[\sum^\infty_{n=1}\frac{1}{n}f_{tot}(e^{ina},
y_{j}^n,x^n)\right].
\end{equation}

 To take trace over Hilbert-space on $S^2$, we
impose proper periodic boundary conditions on time direction
$\mathbb{R}$. As a result, the base manifold becomes $S^1\times S^2$.
For saddle points in localization procedure, we need to turn on
monopole fluxes on $S^2$ and holonomy along $S^1$. These
configurations of the gauge fields are denoted by  $\{m\}$ and $\{ a
\}$ collectively. Both variables take values in the Cartan
subalgebra of $G$. $S_0$ denotes the classical action for the
(monopole+holnomoy) configuration on $S^1\times S^2$. $\epsilon_0$
is called the Casmir energy.
 If the
action contains the Chern-Simons terms, it gives the nonvanishing
contribution,
\begin{equation}
S_0=\frac{i k}{4\pi}\int \textrm{tr}(A_0\wedge
dA_0-\frac{2i}{3}A_0\wedge A_0 \wedge A_0)=i k \textrm{tr} (m \, a)
\end{equation}
where $k$ is the Chern-Simons level. In \eqref{index}, $\sum_{m}$ is
over all integral magnetic monopoles charges,
$f_{tot}=f_{chiral}+f_{vector}$ and $(symmetry) = (\text{the order
of the Weyl group})$.  Each component in \eqref{index} is given by
\begin{eqnarray}
&&S^{(0)}_{CS} = i \sum_{\rho\in R_\Phi} k \rho(m) \rho(a) ,\nonumber \\
&&b_0(a)=-\frac{1}{2}\sum_\Phi\sum_{\rho\in R_\Phi}|\rho(m)|\rho(a),\nonumber\\
&&y_{j}^{q_{0j}} = y_{i}^{\frac{1}{2} \sum_\Phi \sum_{\rho\in R_\Phi} |\rho(m)| F_i (\Phi)}, \nonumber \\
&&\epsilon_0 = \frac{1}{2} \sum_\Phi (1-\Delta_\Phi) \sum_{\rho\in R_\Phi} |\rho(m)|
- \frac{1}{2} \sum_{\alpha \in G} |\alpha(m)|, \nonumber\\
&&f_{chiral}(e^{ia}, y_{j},x) = \sum_\Phi \sum_{\rho\in R_\Phi}
\left[ e^{i\rho(a)} y_{j}^{F_{j}}
\frac{x^{|\rho(m)|+\Delta_\Phi}}{1-x^2}  -  e^{-i\rho(a)}
y_{j}^{-F_{j}} \frac{x^{|\rho(m)|+2-\Delta_\Phi}}{1-x^2} \right]\label{universal}
\end{eqnarray}
where $\sum_\Phi$, $\sum_{\rho\in R_\Phi}$ and $\sum_{\alpha\in G}$  represent the
summations over all chiral multiplets, all weights and all roots, respectively.
$F_i$ are the Cartan generators acting only on the $i$-th Flavor.
In addition, $\textrm{exp}\left[\sum^\infty_{n=1}\frac{1}{n}f_{vector}(e^{ina},x^n)\right]$
can be simplified as follows:
\begin{eqnarray}
\exp\left[\sum^\infty_{n=1}\frac{1}{n}f_{vector}(e^{ina},x^n)\right]&=&\prod_{\alpha\in G}\exp\left[-\sum^\infty_{n=1}\frac{1}{n}e^{in\alpha(a)}x^{n|\alpha(m)|}\right]\nonumber\\
&=&\prod_{\alpha\in G}\exp\left[\ln\left(1-e^{i\alpha(a)}x^{|\alpha(m)|}\right)\right]\nonumber\\
&=&\prod_{\alpha\in G}\left(1-e^{i\alpha(a)}x^{|\alpha(m)|}\right).
\end{eqnarray}

\subsection{Unitary Case}

We consider $\mathcal N=2$ $U(N_c)$ gauge theory with $N_f$ (anti)fundamental
chiral muliplets $Q^a,\tilde{Q}_b$ and a Chern-Simons term at level
$k$. It's magnetic dual is given by $\mathcal N=2$  $U(|k| + N_f - N_c)$
gauge theory with $N_f$ (anti)fundamental chiral multiplets $q_a,
\tilde{q}^b$ and $N_f \times N_f$ matrix of singlets $M^{a}_{b}$
with Chern-Simons term at level $-k$ and the superpotential
\begin{equation}
W=M^{a}_{b}q_a\tilde{q}^b.
\end{equation}
The weights of the fundamental representation are $ \epsilon_i$
where $i=1,\cdots,N_c$, and the roots of $U(N_c)$ are
$\epsilon_i - \epsilon_j$ where $i,j=1,\cdots,N_c$ and $i\neq j$.
The superconformal index without the chemical potentials($y_{j} =
1$) is thus given by:
\begin{eqnarray}
&&S^{(0)}_{CS} = ik \sum^{N_c}_{i=1} a_i m_i,\\
&&b_0(a)=0,\\
&&\epsilon_0 = \left\{ \begin{array}{ll}
\displaystyle N_f (1-r) \sum^{N_c}_{i=1}|m_i| -\sum^{N_c}_{i<j}|m_i-m_j|, &\textrm{Electric}\\
\displaystyle N_f     r \sum^{N_c}_{i=1}|m_i| -\sum^{N_c}_{i<j}|m_i-m_j|, &\textrm{Magnetic}\\
\end{array} \right.\\
&&f_{chiral}(e^{ia},1,x) = \left\{ \begin{array}{ll}
\displaystyle N_f  \frac{ x^{r} - x^{2-r}}{ 1-x^2}   \left[ \sum^{N_c}_{i=1}  x^{|m_i|} 2\cos a_i  \right], &\textrm{Electric}\\
\displaystyle N_f  \frac{ x^{1-r} - x^{1+r}}{ 1-x^2} \left[ \sum^{N_c}_{i=1}  x^{|m_i|} 2\cos a_i  \right] + {N_f}^2 \frac{ x^{2r}-  x^{2-2r}}{ 1-x^2} , &\textrm{Magnetic}\\
\end{array} \right.
\end{eqnarray}
\begin{eqnarray}
\exp \left[ \sum^\infty_{n=1} \frac{1}{n} f_{vector}(e^{ina},x^n) \right]  &=&  \prod^{N_c}_{i<j}\left(1-e^{i(a_i-a_j)}x^{|m_i-m_j|}\right) \left(1-e^{-i(a_i-a_j)}x^{|m_i-m_j|} \right)  \nonumber\\
&=&\prod^{N_c}_{i<j}\left( 1-2\cos(a_i-a_j)x^{|m_i-m_j|}+x^{2|m_i-m_j|} \right).
\end{eqnarray}
Due to the flavor symmetry, one can assume that $Q^a, \tilde{Q}_b$
have the same R-charge $r$. Since $M^{a}_{b}$ is quadratic in $Q,
\tilde{Q}$ it has the R-charge $2r$. Since the superpotential has
the dimension 2 in the IR limit, $q_a, \tilde{q}^b$ has the R-charge
$1-r$.

The index formula can be expanded order by order in terms of variables $p$ and $q$ that are defined by
\begin{equation}\label{pq}
p=x^r,~~~~~~q=x^{1-r}.
\end{equation}
The $r$ dependence of the index can be restored by replacing $p$ and
$q$ in the index formula expanded in terms of $p$ and $q$ by
$\eqref{pq}$. We computed the indices of all possible dual pairs
between the electric theory and the magnetic theory in the range
$1\leq N_c,|k|+N_f-N_c\leq2$ with unfixed R-charge $r$, and
confirmed the agreements up to at least $\mathcal O(p^{12})$ and
$\mathcal O(q^{12})$. We list a part of the result in the following
table:
\begin{center}
\begin{tabular}{|c|c|c|p{7cm}|}
\hline
              &  Electric  &    Magnetic     & \\
$(N_f,k,N_c)$ &  $U(N_c)$  &$U(|k|+N_f-N_c)$ & Index (r is R-charge) \\
\hline
(1,1,1)       &   $U(1)$   &  $U(1)$         & $ 1-x^4-2 x^8+x^{2 r}+x^{4 r}+x^{6 r}+x^{8 r}+x^{-2 r} \left(-x^4-x^8\right)+\cdots $ \\
\hline
(1,2,1)       &   $U(1)$   &  $U(2)$         & $ 1-2 x^2-3 x^4-2 x^{5-r}+x^{4 r}+x^{6 r}+x^{2 r} \left(1-2 x^4\right)+x^r \left(2 x^3+2 x^5\right)+x^{-2 r} \left(x^4+2 x^6\right)+\cdots $ \\
\hline
(2,1,1)       &   $U(1)$   &  $U(2)$         & $1-8 x^2+6 x^4+48 x^6-4 x^{5-3 r}+4 x^{4-2 r}+16 x^{6 r}+12 x^{5+r}+x^{4 r} \left(9-24 x^2\right)+x^{2 r} \left(4-16 x^2-16 x^4\right)+x^{-r} \left(4x^3+4 x^5\right)+\cdots $ \\
\hline
(1,2,2)       &   $U(2)$   &  $U(1)$         & $1-2 x^2-3 x^4-2 x^{6-3 r}+x^{4 r}+x^{6 r}+x^{2 r} \left(1-2 x^4\right)+x^{-2 r} \left(x^4+2 x^6\right)+x^{-r} \left(2 x^4+2 x^6\right)+\cdots$ \\
\hline
(2,1,2)       &   $U(2)$   &  $U(1)$         & $1-8 x^2+28 x^4+32 x^6+24 x^{6-3 r}+20 x^{6 r}+12 x^{2+3 r}+x^{4 r} \left(10-48 x^2\right)+x^r \left(4 x^2-44 x^4\right)+x^{2 r} \left(4-24 x^2+32x^4\right)+x^{-2 r} \left(8 x^4-24 x^6\right)+x^{-4 r} \left(-x^4-8 x^6\right)+x^{-r} \left(-16 x^4+52 x^6\right)+\cdots $ \\
\hline
(1,3,2)       &   $U(2)$   &  $U(2)$         & $ 1-2 x^2-2 x^4+x^{4-2 r}+x^{4 r}+x^{2 r} \left(1-3 x^4\right)+\cdots $ \\
\hline
(2,2,2)       &   $U(2)$   &  $U(2)$         & $ 1-8 x^2-2 x^3+28 x^4+4 x^{4-2 r}+10 x^{4 r}+x^{2 r} \left(4-24 x^2\right)+\cdots $ \\
\hline
(3,1,2)       &   $U(2)$   &  $U(2)$         & $1-18 x^2+18 x^3+198 x^4+9 x^{4-2 r}+45 x^{4 r}+x^{2 r} \left(9-144 x^2\right)+\cdots $ \\
\hline
\end{tabular}
\end{center}

Note that the index matches for arbitrary assignment of the R-charge
for $Q, \tilde{Q}$. To determine the precise value of $r$ we have to
use the other method such as Z-maximization proposed by
\cite{Jafferis10}.

 It's worthwhile to work out the gauge invariant operators
of the first few lowest orders. We are working on $U(N)$ case but
similar argument can be given to other gauge groups. The easiest one
is the chiral ring elements. For $U(N)$ with $N_f$ flavors, it is
given by $Q^a_i\tilde Q^i_b$ where $i$ is a gauge index running from
1 to $N_c$ and $a$, $b$ are flavor indices running from 1 to $N_f$.
The total number of the chiral primaries is $N_f^2$. In the magnetic
side, these are simply given by  $M^a_b$. Due to the superpotential
terms $q_a\tilde q^b$ turn out to be $Q$ exact operators. The chiral
ring elements contribute $+N_f^2 x^{2r}$ to the index. There are
terms
 in the index which do not depend on R-charges such as $x^2, x^4
\cdots$. For lowest such term one can consider the operators
involving fermions.  The fermion operator $\psi^\dagger$ has
R-charge $1-r$ and the spin $\frac{1}{2}$ as the lowest one. Thus,
it gives the contribution of $x^{R+2j}=x^{2-r}$. For $U(N)$ case, we
have  $Q^a\psi^\dagger_b$ or $\tilde Q_a \tilde\psi^{\dagger b}$
terms and each of which contributes $(x^r)(-x^{2-r})$ to the index.
So the index get the contribution $-2N_f^2 x^2$. This explains the
index for the gauge group $U(2)$ and higher rank but for $U(1)$ with
$k=N_f=1$ such term is missing. Thus we have to look for additional
operators. For that purpose, one can consider monopole operators.
For simplicity we consider $U(1)$. One can consider the general
$U(N)$ but the resulting monopole operators will contribute to
higher orders. We use the operator-state correspondence for
conformal field theory and work out states on $S^2 \times R$.  If we
turn on the monopole flux $n$ we have  nonzero matter fields due to
the Gauss constraints. The BPS state can be represented as
\begin{equation}
\left|\tilde Q_{a_1}\tilde Q_{a_2}\cdots\tilde Q_{a_{k|n|}}\right>.
\end{equation}
For each $\tilde Q_{a_i}$ it has the R-charge $r$ and the angular
momentum $\frac{n}{2}$. This is due to the familiar fact that the
charged scalar of charge $e$ has the angular momentum $|e n|$ in the
presence of the monopole charge $n$ on $S^2$. Thus for each
$Q^{a_i}$ we have $\epsilon=R+j=r+\frac{n}{2}$. We can count the
number of such operators by the combination with repetition:
${}_{N_f}H_{kn}=\binom{N_f+kn-1}{kn}=\frac{(N_f+kn-1)!}{(N_f-1)!(kn)!}$.
In addition, if the magnetic flux is negative, we have following
gauge invariant operators in the same manner,
\begin{equation}
\left|Q^{a_1}Q^{a_2}\cdots Q^{a_{kn}}\right>
\end{equation}
Therefore, the contribution of this kind of operators to the
superconformal index is given by
\begin{equation}
\frac{(N_f+k|n|-1)!}{(N_f-1)!(k|n|)!}x^{k|n|^2+N_f|n|+(k-N_f)|n|r}.
\end{equation}
where the power of $x$ is given by
$\epsilon_0+\epsilon+j=N_f(1-r)|n|+k|n|(r+2\times\frac{|n|}{2})=k|n|^2+N_f|n|+(k-N_f)|n|r$.
  For the $N_f=k=1$ case, this contribution becomes $x^{|n|^2+|n|}$;
two terms from $n=1$ and $n=-1$ give $2x^2$, which exactly cancels
the contribution $-2x^2$ from fermionic excitations $Q\psi^\dagger$
and $\tilde Q \tilde\psi^\dagger$. This explains the absence of
$x^2$ term for $U(1)$ gauge group with $N_f=k=1$. Using the chiral
ring elements and the monopole operators discussed above, one can
understand the numerical value of the index of the few lowest orders
in $x$.

\subsection{Symplectic Case}

Now turn to $\mathcal N=2$ $Sp(2N_c)$ gauge theory with $2N_f$
chiral muliplets $Q^a$ and a Chern-Simons term at level $k$. Here
$k$ and $N_f$ may be half-integral, but must sum to an integer. It's
magnetic dual is given by $\mathcal N=2$ $Sp(2(|k|+N_f-N_c-1))$
gauge theory with $2N_f$ chiral multiplets $q_a$ and a Chern-Simons
term at level $-k$. In addition, there are $N_f(2N_f-1)$ uncharged
chiral multiplets $M^{ab}$, which couple through a superpotential
that is given by
\begin{equation}
W=M^{ab} q_a q_b.
\end{equation}
The weights of the fundamental representation are $\pm \epsilon_i$
where $i=1,\cdots,N_c$, and the roots of $Sp(2N_c)$ are $\pm
2\epsilon_i$ and $\pm\epsilon_i\pm\epsilon_j$ where
$i,j=1,\cdots,N_c$ and $i\neq j$. The computation is straightforward
and we just list the results. We computed the indices of all dual
pairs in the range $1\leq N_c,|k|+N_f-N_c-1\leq2$ with unfixed
R-charge $r$, and confirmed the agreements up to at least $\mathcal
O(p^{12})$ and $\mathcal O(q^{12})$. Parts of them are listed in the
following table:
\begin{center}
\begin{tabular}{|c|c|c|p{5.5cm}|}
\hline
              & Electric    &   Magnetic                     &  \\
$(N_f,k,N_c)$ & $Sp(2N_c)$  &   $Sp(2(|k| + N_f - N_c - 1)$  & Index (r is R-charge) \\
\hline
(1,2,1)       &   $Sp(2)$   &  $Sp(2)$         &    $ 1-4 x^2-5 x^4+4 x^6+14x^8-x^{8-2r}+x^{4 r}+x^{6 r}+x^{8 r}+x^{2 r} \left(1-4 x^4\right)+x^{-2 r} \left(3 x^4+4 x^6\right)+\cdots $\\
\hline
(1,3,1)       &   $Sp(2)$   &  $Sp(4)$         &    $ 1-4 x^2-2 x^4+16 x^6+3 x^{4-2 r}+x^{4 r}+x^{6 r}+x^{2 r} \left(1-9 x^4\right)+\cdots $\\
\hline
(2,2,1)       &   $Sp(2)$   &  $Sp(4)$         &    $ 1-16 x^2+88 x^4+19 x^6+50 x^{6 r}+x^{4 r} \left(20-160 x^2\right)+x^{2 r} \left(6-64 x^2+156 x^4\right)+x^{-2 r} \left(10 x^4-74 x^6\right)+\cdots $\\
\hline
(1,3,2)       &   $Sp(4)$   &  $Sp(2)$         &    $ 1-4 x^2-2 x^4+12 x^6+x^{4 r}+x^{6 r}+x^{2 r} \left(1-5 x^4\right)+x^{-2 r} \left(3 x^4-4 x^6\right)+\cdots $\\
\hline
(1,4,2)       &   $Sp(4)$   &  $Sp(4)$         &    $ 1-4 x^2+x^4+3x^{4-2 r}+x^{4 r}+x^{2 r} \left(1-10 x^4\right)+\cdots $ \\
\hline
(2,3,2)       &   $Sp(4)$   &  $Sp(4)$         &    $ 1-16 x^2+148 x^4+10 x^{4-2 r}+21 x^{4 r}+x^{2 r} \left(6-80 x^2\right)+\cdots $ \\
\hline
(3,2,2)       &   $Sp(4)$   &  $Sp(4)$         &    $ 1-36 x^2+873 x^4+21 x^{4-2 r}+120 x^{4 r}+x^{2 r} \left(15-504 x^2\right)+\cdots $ \\
\hline
(4,1,2)       &   $Sp(4)$   &  $Sp(4)$         &    $ 1-64 x^2+2896 x^4+36 x^{4-2 r}+406 x^{4 r}+x^{2 r} \left(28-1728 x^2\right)+\cdots  $ \\
\hline
\end{tabular}
\end{center}

\subsection{Orthogonal Case}

The electric theory is given by $\mathcal N=2$ $O(N_c)$ gauge theory with $N_f$
flavors of chiral superfields $Q^a$, $a=1,\cdots,N_f$ in the vector
representation and no superpotential. Its magnetic dual is given by
$O(N_f-N_c+|k|+2)$ gauge theory with $N_f$ flavors of chiral
superfields $q_a$ in the vector representation as well as a singlet
chiral superfield $M^{ab}$ which is a symmetric $N_f\times N_f$
matrix. The superpotential in the magnetic theory is
\begin{equation}
W=M^{ab} q_a q_b.
\end{equation}
Let us first consider $O(2N)$ case. The index formula is given by
\eqref{index}. With facts that the weights of the fundamental representation
are $\pm \epsilon_i$ where $i=1,\cdots,N$ and that the roots of
$O(2N)$ are $\pm\epsilon_i\pm\epsilon_j$ where $i,j=1,\cdots,N$ and
$i\neq j$,
\begin{eqnarray}
&&S^{(0)}_{CS}= i\frac{k}{2}\sum^{N}_{i=1}2a_im_i =ik\sum^{N}_{i=1}a_im_i,\\
&&b_0(a)=0,\\
&&\epsilon_0 = \left\{ \begin{array}{ll}
\displaystyle N_f(1-r)\sum^{N}_{i=1}|m_i|-\sum^{N}_{i<j}|m_i+m_j|-\sum^{N}_{i<j}|m_i-m_j|, &\textrm{Electric}\\
\displaystyle N_f r   \sum^{N}_{i=1}|m_i|-\sum^{N}_{i<j}|m_i+m_j|-\sum^{N}_{i<j}|m_i-m_j|, & \textrm{Magnetic}\\
\end{array} \right.\\
&&f_{chiral}(e^{ia}, 1,x) = \left\{ \begin{array}{ll}
\displaystyle N_f  \frac{ x^{r} -  x^{2-r}}{1-x^2}   \left[ \sum^{N}_{i=1}  x^{|m_i|} 2\cos a_i  \right], &\textrm{Electric}\\
\displaystyle N_f  \frac{ x^{1-r} -  x^{1+r}}{1-x^2} \left[ \sum^{N}_{i=1}  x^{|m_i|} 2\cos a_i  \right] \\
\displaystyle ~~ + \frac{N_f(N_f+1)}{2} \frac{ x^{2r}-  x^{2-2r}}{1-x^2} , & \textrm{Magnetic}\\
\end{array} \right.
\end{eqnarray}
\begin{eqnarray}
\exp\left[\sum^\infty_{n=1}\frac{1}{n}f_{vector}(e^{ina},x^n)\right]&=&\prod^{N}_{i<j}\left(1-e^{i(a_i+a_j)}x^{|m_i+m_j|}\right)\left(1-e^{-i(a_i+a_j)}x^{|m_i+m_j|}\right)\nonumber\\
&&~~~~\times\left(1-e^{i(a_i-a_j)}x^{|m_i-m_j|}\right)\left(1-e^{-i(a_i-a_j)}x^{|m_i-m_j|}\right)\nonumber\\
&=&\prod^{N}_{i<j}\left(1-2\cos(a_i+a_j)x^{|m_i+m_j|}+x^{2|m_i+m_j|}\right)\nonumber\\
&&~~~~\times\left(1-2\cos(a_i-a_j)x^{|m_i-m_j|}+x^{2|m_i-m_j|}\right).
\end{eqnarray}
This index formula holds for $SO(2N)$ case. We should consider the
additional projection for $Z_2$ element of $O(2N)$ not belonging to
$SO(2N)$ group. This kind of projection was considered before in the
superconformal index computation for $\mathcal N=5$ super Chern-Simons matter
theories \cite{Cheon} and we adopt the procedure to our purpose. We choose the
specific $Z_2$ action,
\begin{equation}
Z_2= \left(\begin{array}{cccc}
1&&&\\
&-1&&\\
&&1&\\
&&&\ddots
\end{array}\right).
\end{equation}
Under this $Z_2$ action, the eigenvalues of the holonomy and the
monopole are projected into
\begin{equation}
e^{\pm ia_1}\rightarrow\pm1,~~~~~~\pm m_1\rightarrow0.
\end{equation}
The other variables are not affected. Thus, $f_{chiral}$ turns into
\begin{eqnarray}
f_{chiral}(e^{ia}, 1,x) &=& \left\{ \begin{array}{ll}
\displaystyle N_f  \frac{ x^{r} -  x^{2-r}}{1-x^2}   \left[ (1+(-1)^n) + \sum^{N}_{i=2} x^{|m_i|} 2\cos a_i  \right],
 &\textrm{Electric}\\
\displaystyle N_f  \frac{ x^{1-r} - x^{1+r}}{1-x^2} \left[ (1+(-1)^n) + \sum^{N}_{i=2} x^{|m_i|} 2\cos a_i  \right]\\
\displaystyle ~~ + \frac{N_f(N_f+1)}{2} \frac{ x^{2r}- x^{2-2r}}{1-x^2} , & \textrm{Magnetic}\\
\end{array} \right.
\end{eqnarray}
and the vector term changes into
\begin{eqnarray}
\exp\left[\sum^\infty_{n=1}\frac{1}{n}f_{vector}(e^{ina},x^n)\right]&=&\prod^{N}_{i=2}\left(1-2i\sin(a_i)x^{|m_i|}-x^{2|m_i|}\right)\left(1+2i\sin(a_i)x^{|m_i|}-x^{2|m_i|}\right)\nonumber\\
&&\prod^{N}_{1<i<j}\left(1-2\cos(a_i+a_j)x^{|m_i+m_j|}+x^{2|m_i+m_j|}\right)\nonumber\\
&&~~~~\times\left(1-2\cos(a_i-a_j)x^{|m_i-m_j|}+x^{2|m_i-m_j|}\right)
.
\end{eqnarray}
Other terms are obtained simply by setting $m_1=0$.

Let us turn to $O(2N+1)$ theory. With facts that the weights of the fundamental
representation are $\pm \epsilon_i$ where $i=1,\cdots,N$ and that
the roots of $O(2N+1)$ are $\pm \epsilon_i$ and
$\pm\epsilon_i\pm\epsilon_j$ where $i,j=1,\cdots,N$ and $i\neq j$,
\begin{eqnarray}
&&S^{(0)}_{CS} = ik\sum^{N}_{i=1}a_im_i,\\
&&b_0(a) = 0,\\
&&\epsilon_0 = \left\{ \begin{array}{ll}
\displaystyle N_f(1-r)\sum^{N}_{i=1}|m_i|-\sum^{N}_{i=1}|m_i|-\sum^{N}_{i<j}|m_i+m_j|-\sum^{N}_{i<j}|m_i-m_j|, &\textrm{Electric}\\
\displaystyle N_f   r \sum^{N}_{i=1}|m_i|-\sum^{N}_{i=1}|m_i|-\sum^{N}_{i<j}|m_i+m_j|-\sum^{N}_{i<j}|m_i-m_j|, & \textrm{Magnetic}\\
\end{array} \right.\\
&&f_{chiral}(e^{ia},1,x) = \left\{ \begin{array}{ll}
\displaystyle N_f  \frac{ x^{r}  -  x^{2-r}}{1-x^2} \left[ \sum^{N}_{i=1}  x^{|m_i|} 2\cos a_i +1 \right], &\textrm{Electric}\\
\displaystyle N_f  \frac{ x^{1-r} - x^{1+r}}{1-x^2} \left[ \sum^{N}_{i=1} x^{|m_i|} 2\cos a_i +1 \right]\\
\displaystyle ~~ + \frac{N_f(N_f+1)}{2} \frac{ x^{2r}-  x^{2-2r}}{1-x^2}. & \textrm{Magnetic}\\
\end{array} \right.
\end{eqnarray}
Note that we have to understand $e^{i\rho(a)}$ in the chiral letter index as the eigenvalues of the operator $e^{ia}$, which are $e^{\pm a_i}$ and 1 where $i=1,\cdots,N$.\\
In addition,
$\textrm{exp}\left[\sum^\infty_{n=1}\frac{1}{n}f_{vector}(e^{ima},x^m)\right]$
can be simplified as follows:
\begin{eqnarray}
\exp\left[\sum^\infty_{n=1}\frac{1}{n}f_{vector}(e^{ima},x^m)\right]
&=&\prod^{N}_{i=1}\left(1-2\cos a_ix^{|m_i|}+x^{2|m_i|}\right)\nonumber\\
&&\prod^{N}_{i<j}\left(1-2\cos(a_i+a_j)x^{|m_i+m_j|}+x^{2|m_i+m_j|}\right)\nonumber\\
&& ~ ~ ~ ~ \times\left(1-2\cos(a_i-a_j)x^{|m_i-m_j|}+x^{2|m_i-m_j|}\right).\nonumber\\
\end{eqnarray}
Again we have to consider the further projection due to proper $O(2N+1)$
elements. Under the $Z_2$ action,
\begin{equation}
Z_2= \left(\begin{array}{cccc}
1&&&\\
&\ddots&&\\
&&1&\\
&&&-1
\end{array}\right),
\end{equation}
an eigenvalue 1 of the holonomy in the fundamental representation is
projected by
\begin{equation}
1\rightarrow-1
\end{equation}
while the others are not influenced. Furthermore, eigenvalues
$e^{\pm ia_i}$ of the holonomy in the adjoint representation are
projected by
\begin{equation}
e^{\pm ia_i}=e^{\pm ia_i}\cdot1\rightarrow e^{\pm ia_1}\cdot(-1)
\end{equation}
while the others, which are in the form of $e^{i(\pm a_i\pm
a_j)}=e^{\pm ia_i}\cdot e^{\pm ia_i}$, are not influenced. Thus, the
projected index is obtained from
\begin{equation}
f_{chiral}(e^{ia},1,x) = \left\{ \begin{array}{ll}
\displaystyle N_f  \frac{ x^{r}   -  x^{2-r}}{1-x^2} \left[ \sum^{N}_{i=1}  x^{|m_i|} 2\cos a_i +(-1)^n \right], &\textrm{Electric}\\
\displaystyle N_f  \frac{ x^{1-r} -  x^{1+r}}{1-x^2} \left[ \sum^{N}_{i=1}  x^{|m_i|} 2\cos a_i +(-1)^n \right]\\
\displaystyle ~~ + \frac{N_f(N_f+1)}{2}  \frac{ x^{2r}- x^{2-2r}}{1-x^2} , & \textrm{Magnetic}\\
\end{array} \right.
\end{equation}
and
\begin{eqnarray}
\exp\left[\sum^\infty_{n=1}\frac{1}{n}f_{vector}(e^{ina},x^n)\right]&=&\prod^{N}_{i=1}\left(1+2\cos a_ix^{|m_i|}+x^{2|m_i|}\right)\nonumber\\
&&\prod^{N}_{i<j}\left(1-2\cos(a_i+a_j)x^{|m_i+m_j|}+x^{2|m_i+m_j|}\right)\nonumber\\
&& ~ ~ ~ ~ \times\left(1-2\cos(a_i-a_j)x^{|m_i-m_j|}+x^{2|m_i-m_j|}\right).\nonumber\\
\end{eqnarray}

 We computed the indices of all dual pairs in the range $1\leq
N_c,|k|+N_f-N_c+2\leq 4$ with unfixed R-charge $r$, and confirmed the
agreements up to at least $\mathcal O(p^{12})$ and $\mathcal
O(q^{12})$. It is crucial that we have $O(N)$ gauge group instead of
$SO(N)$ to have the agreements in index for the proposed dual pairs.
Parts of them are listed in the following table:
\begin{center}
\begin{tabular}{|c|c|c|p{6cm}|}
\hline
              & Electric    &   Magnetic                    &  \\
$(N_f,k,N_c)$ & $O(N_c)$    &   $O(|k| + N_f - N_c +2)  $    & Index (r is R-charge) \\
\hline
(1,1,1)       &   $O(1)$   &  $O(3)$         &    $ 1-x^2-2 x^4-2 x^6-2 x^8+x^{4 r}+x^{6 r}+x^{8 r}+x^{2 r} \left(1-x^6\right)+x^{-2 r} \left(x^6+x^8\right)+\cdots $\\
\hline
(1,1,2)       &   $O(2)$   &  $O(2)$         &    $ 1-x^4-2 x^8+x^{2 r}+x^{4 r}+x^{6 r}+x^{8 r}+x^{-2 r} \left(-x^4-x^8\right)+\cdots $\\
\hline
(1,2,2)       &   $O(2)$   &  $O(3)$         &    $ 1-x^2-2 x^4+x^{6-2 r}-x^{5-r}+x^{4 r}+x^{6 r}+x^{2 r} \left(1-x^4\right)+x^r \left(x^3+x^5\right)+\cdots $\\
\hline
(1,3,2)       &   $O(2)$   &  $O(4)$         &    $ 1-x^2-2 x^4-x^6+x^{6-2 r}+x^{2 r}+x^{4 r}+x^{6 r}+\cdots $\\
\hline
(1,3,3)       &   $O(3)$   &  $O(3)$         &    $ 1-x^2-2 x^4+x^5+x^{6-2 r}+x^{4 r}+x^{6 r}+x^{2 r} \left(1-x^4\right)+\cdots $\\
\hline
(1,4,3)       &   $O(3)$   &  $O(4)$         &    $ 1-x^2-2 x^4+x^{6-2 r}+x^{4 r}+x^{6 r}+x^{6+r}+x^{2 r} \left(1-x^4-2 x^6\right)+\cdots $\\
\hline
(2,1,4)       &   $O(4)$   &  $O(1)$         &    $ 1-4 x^2+10x^4-x^{4-4 r}+2x^{4-2 r}+x^{4 r} \left(6-12 x^2-19 x^4\right)+x^{2 r} \left(3-8 x^2-9 x^4\right)+\cdots $\\
\hline
(5,1,4)       &   $O(4)$   &  $O(4)$         &    $ 1-25 x^2+475 x^4+10 x^{4-2 r}+120 x^{4 r}+x^{2 r} \left(15-350 x^2\right)+\cdots $\\
\hline
\end{tabular}
\end{center}

\section{Conclusions}
We work out the superconformal index for Seiberg-like dual pairs in
three-dimensional Chern-Simons matter theories with gauge group
$U(N)/Sp(2N)/O(N)$ with matters with $N, 2N, N$-dimensional
representation, respectively. We find perfect agreements as far as
we can carry out the numerical computation. It would be interesting
to attempt the analytic proof for the equality of the index for the
dual pairs. Related discussion appears at \cite{Vartanov, Rastelli}. 
Certainly the method adopted in the current work is
applicable to other dualities. It would be interesting to carry out
the similar index computation for various proposed dual pairs. The
index computation is a useful tool to confirm proposed dualities as
demonstrated in the current work.

\vskip 0.5cm  \hspace*{-0.8cm} {\bf\large Acknowledgements} \vskip
0.2cm

\hspace*{-0.75cm} We are grateful to Dongmin Gang for the discussion
on the computation of the superconformal index. J.P.  is supported
by the KOSEF Grant R01-2008-000-20370-0, the National Research
Foundation of Korea (NRF) Grants No. 2009-0085995  and 2005-0049409
through the Center for Quantum Spacetime (CQUeST) of Sogang
University. J. P. also appreciates APCTP for its stimulating
environment for research.

\newpage
\appendix
\section{Appendix: Index computation with chemical potentials}
In appendix, we list the results of index computations,
turning on the chemical potentials for the flavor symmetry. When the
chemical potentials are turned on, only the flavor charge terms
$y_{j}^{q_{0j}}$ and the chiral letter index $f_{chiral}$ are
different from those in the main text. These can be read off from
the universal formula eq. (\ref{universal})

\subsection{Unitary Case} Besides the R symmetry, the (global) symmetries of the unitary case are given
by $U(N_f) \times U(N_f)=U(1)_A \times U(1)_B \times SU(N_f) \times
SU(N_f)$. We introduce the corresponding Cartan generators $F_{i}$
and $G_{j}$ for which the charge assignments of the matter contents
on each of the electric side and the magnetic side are as follows
\begin{equation}
F_{i}(Q^a)=F_{i}(\tilde Q_a)=\delta_{ia},~~~~~~
G_{j}(Q^a)=-G_{j}(\tilde Q_a)=\delta_{ja},
\end{equation}
\begin{equation}
F_{i}(q_a)=F_{i}(\tilde q^a)=-\delta_{ia},~~~~~~F_{i}(M^a_b)=\delta_{ia}+\delta_{ib},
\end{equation}
\begin{equation}
G_{j}(q_a)=-G_{j}(\tilde q^a)=\delta_{ja},~~~~~~G_{j}(M^a_b)=\delta_{ja}-\delta_{jb}
\end{equation}
where $i,j=1,2,\cdots,N_f$. $U(1)_A$ and  $U(1)_B$ are generated by
$\Sigma_{i=1}^{N_f}F_i$ and $\Sigma_{j=1}^{N_f}G_j$ respectively.
Note that $U(1)_B$ is a gauged symmetry. The operators $y_i^{F_{i}}$
and $z_j^{G_{j}}$ then contribute to the superconformal index as
follows:
\begin{eqnarray}
y_{j}^{q_{0j}} &=& y_{i}^{\frac{1}{2} \sum_\Phi \sum_{\rho\in R_\Phi} |\rho(m)| F_{i}(\Phi)}
\times z_{j}^{\frac{1}{2} \sum_\Phi \sum_{\rho\in R_\Phi} |\rho(m)| G_{j}(\Phi)}\nonumber\\
&=&\left\{ \begin{array}{ll}
\displaystyle \prod^{N_f}_{j=1} y_j^{ - \frac{1}{2} \sum^{N_c}_{i=1}|m_i| (1+1) }
\times z_j^{ - \frac{1}{2} \sum^{N_c}_{i=1}|m_i| (1-1) } = \prod^{N_f}_{j=1}
y_j^{ - \sum^{N_c}_{i=1}|m_i| }, & \textrm{Electric}\\
\displaystyle \prod^{N_f}_{j=1} y_j^{ - \frac{1}{2} \sum^{N_c}_{i=1}|m_i| (-1-1)}
\times z_j^{ - \frac{1}{2} \sum^{N_c}_{i=1}|m_i| (1-1) } = \prod^{N_f}_{j=1} y_j^{
\sum^{N_c}_{i=1}|m_i| }, & \textrm{Magnetic}\\
\end{array} \right.
\end{eqnarray}
\begin{equation}
f_{chiral}(e^{ia}, y_{j}z_{j},x) = \left\{ \begin{array}{ll}
\displaystyle \sum^{N_f}_{j=1}  \frac{y_j^{1} x^{r} - y^{-1}_j x^{2-r}}{1-x^2}
 \left[ \sum^{N_c}_{i=1}  x^{|m_i|} (z_j^{1} e^{i a_i} - z_j^{-1} e^{- i a_i} )  \right], &\textrm{Electric}\\
\displaystyle \sum^{N_f}_{j=1}  \frac{y_j^{-1} x^{1-r} - y^{1}_j x^{1+r}}{1-x^2}
\left[ \sum^{N_c}_{i=1}  x^{|m_i|} (z_j^{1} e^{i a_i} - z_j^{-1} e^{- i a_i} )  \right]\\
\displaystyle ~~ + \sum^{N_f}_{i=1}  \sum^{N_f}_{j=1}
\frac{y_i^{1} y_j^{1} z_i^{1} z_j^{-1} x^{2r}- y_i^{-1} y_j^{-1} z_i^{-1} z_j^{1} x^{2-2r}}{1-x^2}. &\textrm{Magnetic}\\
\end{array} \right.
\end{equation}

We checked again every case discussed in the main text, turning on the chemical potentials.
 Here, we simply give one example since writing down the full results is
 rather cumbersome. For $(N_f,k,N_c) = (2,1,1)$, the electric $U(1)$ and the magnetic $U(2)$,
\begin{eqnarray}
&I&(x,y_1,y_2,z_1,z_2)\nonumber\\
&=&1+x^2 \left(-4-\frac{y_1 z_1}{y_2 z_2}-\frac{y_2 z_1}{y_1 z_2}-\frac{y_1 z_2}{y_2 z_1}
-\frac{y_2 z_2}{y_1 z_1}\right)\nonumber\\
&&+x^4\left(-2+\frac{y_1^2}{y_2^2}+\frac{y_2^2}{y_1^2}+\frac{z_1^2}{z_2^2}
+\frac{y_1 z_1}{y_2 z_2}+\frac{y_2 z_1}{y_1 z_2}+\frac{y_1 z_2}{y_2 z_1}
+\frac{y_2z_2}{y_1 z_1}+\frac{z_2^2}{z_1^2}\right)\nonumber\\
&&+x^{4-2 r} \left(\frac{1}{y_1^2}+\frac{1}{y_2^2}+\frac{z_1}{y_1y_2 z_2}+\frac{z_2}{y_1 y_2 z_1}\right)
+x^{3-r} \left(\frac{1}{y_2 z_1}+\frac{z_1}{y_2}+\frac{1}{y_1 z_2}+\frac{z_2}{y_1}\right)\nonumber\\
&&+x^{2 r} \left(y_1^2+y_2^2+\frac{y_1 y_2z_1}{z_2}+\frac{y_1 y_2 z_2}{z_1}
+x^2 \left(-2 y_1^2-2 y_2^2-\frac{y_1^2 z_1^2}{z_2^2}-\frac{y_2^2 z_1^2}{z_2^2}-\frac{y_1^3 z_1}{y_2 z_2}\right.\right.\nonumber\\
&&~~~~\left.\left.-\frac{2y_1 y_2 z_1}{z_2}-\frac{y_2^3 z_1}{y_1 z_2}
-\frac{y_1^3 z_2}{y_2 z_1}-\frac{2 y_1 y_2 z_2}{z_1}-\frac{y_2^3 z_2}{y_1 z_1}-\frac{y_1^2 z_2^2}{z_1^2}
-\frac{y_2^2z_2^2}{z_1^2}\right)\right)\nonumber\\
&&+x^{4 r} \left(y_1^4+y_1^2 y_2^2+y_2^4+\frac{y_1^2 y_2^2 z_1^2}{z_2^2}+\frac{y_1^3 y_2 z_1}{z_2}
+\frac{y_1y_2^3 z_1}{z_2}+\frac{y_1^3 y_2 z_2}{z_1}+\frac{y_1 y_2^3 z_2}{z_1}
+\frac{y_1^2 y_2^2 z_2^2}{z_1^2}\right)\nonumber\\
&&+\cdots\nonumber\\
&=&1-x^2\left(\chi_1(u)+\chi_1(v)+2\big)+x^4\big(\chi_1(u)\chi_1(v)-3\right)\nonumber\\
&&+x^{4-2r}y_0^{-2}\chi_\frac{1}{2}(u)\chi_\frac{1}{2}(v)+x^{3-r}y_0^{-1}\left(z_0\chi_\frac{1}{2}(u)
+z_0^{-1}\chi_\frac{1}{2}(v)\right)\nonumber\\
&&+x^{2r}y_0^2\chi_\frac{1}{2}(u)\chi_\frac{1}{2}(v)\left(1+x^2\left(\chi_1(u)+\chi_1(v)-2\right)\right)\nonumber\\
&&+x^{4r}y_0^4\left(\chi_1(u)\chi_1(v)-1\right)+\cdots
\end{eqnarray}
where $\chi_n(u)=u^{-n}+u^{-n+1}+\cdots+u^n$. $\chi_n(u)$ is the
character of $SU(2)$. A set of variables $y_0=(y_1y_2)^{1/2}$,
$z_0=(z_1z_2)^{1/2}$, $u=\frac{y_1z_1}{y_2z_2}$ and
$v=\frac{y_1z_2}{y_2z_1}$ correspond to the chemical potentials for
 the  symmetries $U(1)_A \times U(1)_B \times SU(2)_Q \times SU(2)_{\tilde Q}$.

\subsection{Symplectic Case} Besides the R symmtery, the global
symmetries of the symplectic case are $U(2N_f)=U(1)_A\times SU(2N_f)$. We introduce the Cartan generators
$F_{i}$ for which the charge assignments of the matter contents are
the followings:
\begin{equation}
F_{i}(Q^a)=\delta_{ia},~~~~~~F_{i}(q_a)=-\delta_{ia},~~~~~~F_{i}(M^{ab})=\delta_{ia}+\delta_{ib}
\end{equation}
where $i=1,2,\cdots,2N_f$. $U(1)_A$ is generated by $\Sigma_{i=1}^{2N_f}F_i$. The operators $y_i^{F_{i}}$ then contribute to the index as follows:
\begin{eqnarray}
&&y_{j}^{q_{0j}} = \left\{ \begin{array}{ll}
\displaystyle \prod^{2N_f}_{j=1} y_j^{ - \frac{1}{2} \sum^{N_c}_{i=1}|2m_i| (1) }
= \prod^{2N_f}_{j=1} y_j^{ - \sum^{N_c}_{i=1}|m_i| }, & \textrm{Electric}\\
\displaystyle \prod^{2N_f}_{j=1} y_j^{ - \frac{1}{2} \sum^{N_c}_{i=1}|2m_i| (-1)}
= \prod^{2N_f}_{j=1} y_j^{   \sum^{N_c}_{i=1}|m_i| }, & \textrm{Magnetic}\\
\end{array} \right.\\
&&f_{chiral}(e^{ia}, y_{j},x) = \left\{ \begin{array}{ll}
\displaystyle \sum^{2N_f}_{j=1}  \frac{y_j^{1} x^{r} - y^{-1}_j x^{2-r}}{1-x^2}
\left[ \sum^{N_c}_{i=1}  x^{|m_i|} 2\cos a_i  \right], &\textrm{Electric}\\
\displaystyle \sum^{2N_f}_{j=1}  \frac{y_j^{-1} x^{1-r} - y^{1}_j x^{1+r}}{1-x^2}
\left[ \sum^{N_c}_{i=1}  x^{|m_i|} 2\cos a_i  \right]\\
\displaystyle ~~ + \sum^{2N_f}_{i=1}  \sum^{2N_f}_{j=i+1} \frac{y_i^{1} y_j^{1} x^{2r}
- y_i^{-1} y_j^{-1} x^{2-2r}}{1-x^2}. & \textrm{Magnetic}\\
\end{array} \right.
\end{eqnarray}

We checked every case discussed in the main text, turning on the chemical potentials,
and give one example: $(N_f,k,N_c) = (1,3,1)$, the electric $Sp(2)$ and
the magnetic $Sp(4)$,
\begin{eqnarray}
&I&(x,y_1,y_2)\nonumber\\
&=&  1+x^2 \left(-2-\frac{y_1}{y_2}-\frac{y_2}{y_1}\right)
+x^6\left(4+\frac{2 y_1^2}{y_2^2}+\frac{4 y_1}{y_2}+\frac{4 y_2}{y_1}+\frac{2 y_2^2}{y_1^2}\right)\nonumber \\
&&+x^{4-2 r}\left(1+\frac{1}{y_1^2}+\frac{1}{y_1}\right)+x^{2 r}
\left(y_1 y_2+x^4 \left(-2 y_1^2-\frac{y_1^3}{y_2}-3y_1 y_2-2 y_2^2-\frac{y_2^3}{y_1}\right)\right)\nonumber \\
&&+x^{4 r} y_1^2 y_2^2+x^{6 r} y_1^3 y_2^3 +\cdots\nonumber \\
&=&1-x^2\left(\chi_1(u)+1\right)-2x^4+2x^6\left(\chi_2(u)+\chi_1(u)\right)+ x^{4-2r}y_0^{-2}\chi_1(u) \nonumber \\
&& +x^{2r}y_0^2\left(1-x^4\left(\chi_2(u)+\chi_1(u)+1\right)\right) +x^{4r}y_0^4+x^{6r}y_0^6+\cdots
\end{eqnarray}
where $y_0=(y_1y_2)^{1/2}$, $u=\frac{y_1}{y_2}$ correspond to the
chemical potentials for global symmetries $U(1)_A \times SU(2)$
respectively.

\subsection{Orthogonal Case} Besides the R symmtery, the global
symmetries of the orthogonal case are given by $U(N_f)=U(1)_A\times SU(N_f)$. We introduce the Cartan generators
$F_{i}$ for which the charge assignments of the matter contents are
the followings:
\begin{equation}
F_{i}(Q^a)=\delta_{ia},~~~~~~F_{i}(q_a)=-\delta_{ia},~~~~~~F_{i}(M^{ab})=\delta_{ia}+\delta_{ib}
\end{equation}
where $i=1,2,\cdots,N_f$. $U(1)_A$ is generated by $\Sigma_{i=1}^{N_f}F_i$.

\subsubsection{O(2N) Theory}
 The operators
$y_i^{F_{i}}$ contribute to the index as follows:
\begin{eqnarray}
&&y_{j}^{q_{0j}} = \left\{ \begin{array}{ll}
\displaystyle \prod_{j=1}^{N_f} y_j^{ - \frac{1}{2} \sum^{N}_{i=1}|2m_i| (1) }
= \prod^{N_f}_{j=1} y_j^{ - \sum^{N}_{i=1}|m_i| }, & \textrm{Electric}\\
\displaystyle \prod_{j=1}^{N_f} y_j^{ - \frac{1}{2} \sum^{N}_{i=1}|2m_i| (-1)}
= \prod^{N_f}_{j=1} y_j^{   \sum^{N}_{i=1}|m_i| }, & \textrm{Magnetic}\\
\end{array} \right.\\
&&f_{chiral}(e^{ia}, y_{j},x) = \left\{ \begin{array}{ll}
\displaystyle \sum^{N_f}_{j=1}  \frac{y_j^{1} x^{r} - y^{-1}_j x^{2-r}}{1-x^2}
 \left[ \sum^{N}_{i=1}  x^{|m_i|} 2\cos a_i  \right], &\textrm{Electric}\\
\displaystyle \sum^{N_f}_{j=1}  \frac{y_j^{-1} x^{1-r} - y^{1}_j x^{1+r}}{1-x^2}
\left[ \sum^{N}_{i=1}  x^{|m_i|} 2\cos a_i  \right]\\
\displaystyle ~~ + \sum^{N_f}_{i=1}  \sum^{N_f}_{j=i} \frac{y_i^{1} y_j^{1} x^{2r}
- y_i^{-1} y_j^{-1} x^{2-2r}}{1-x^2}. & \textrm{Magnetic}\\
\end{array} \right.
\end{eqnarray}
By the projection, the chiral letter index changes into
\begin{equation}
f_{chiral}(e^{ia}, y_{j},x) = \left\{ \begin{array}{ll}
\displaystyle \sum^{N_f}_{j=1}  \frac{y_j^{1} x^{r} - y^{-1}_j x^{2-r}}{1-x^2}   \left[ (1+(-1)^n)
 + \sum^{N}_{i=2} x^{|m_i|} 2\cos a_i  \right], &\textrm{Electric}\\
\displaystyle \sum^{N_f}_{j=1}  \frac{y_j^{-1} x^{1-r} - y^{1}_j x^{1+r}}{1-x^2} \left[ (1+(-1)^n)
 + \sum^{N}_{i=2} x^{|m_i|} 2\cos a_i  \right]\\
\displaystyle ~~ + \sum^{N_f}_{i=1}  \sum^{N_f}_{j=i} \frac{y_i^{1} y_j^{1} x^{2r}
- y_i^{-1} y_j^{-1} x^{2-2r}}{1-x^2}. & \textrm{Magnetic}\\
\end{array} \right.
\end{equation}

\subsubsection{$O(2N+1)$ Theory}
The operators $y_i^{F_{i}}$ contribute to the index as follows:
\begin{eqnarray}
&&y_{j}^{q_{0j}} = \left\{ \begin{array}{ll}
\displaystyle \prod_{j=1}^{N_f} y_j^{ - \frac{1}{2} \sum^{N}_{i=1}|2m_i| (1) } =
\prod^{N_f}_{j=1} y_j^{ - \sum^{N}_{i=1}|m_i| }, & \textrm{Electric}\\
\displaystyle \prod_{j=1}^{N_f} y_j^{ - \frac{1}{2} \sum^{N}_{i=1}|2m_i| (-1)} =
\prod^{N_f}_{j=1} y_j^{   \sum^{N}_{i=1}|m_i| }, & \textrm{Magnetic}\\
\end{array} \right.\\
&&f_{chiral}(e^{ia},y_{j},x) = \left\{ \begin{array}{ll}
\displaystyle \sum^{N_f}_{j=1}  \frac{y_j^{1} x^{r}    - y^{-1}_j x^{2-r}}{1-x^2}
\left[ \sum^{N}_{i=1}  x^{|m_i|} 2\cos a_i +1 \right], &\textrm{Electric}\\
\displaystyle \sum^{N_f}_{j=1}  \frac{y_j^{-1} x^{1-r} - y^{ 1}_j x^{1+r}}{1-x^2}
 \left[ \sum^{N}_{i=1} x^{|m_i|} 2\cos a_i +1 \right]\\
\displaystyle ~~ + \sum^{N_f}_{i=1}  \sum^{N_f}_{j=i} \frac{y_i^{1} y_j^{1} x^{2r}
- y_i^{-1} y_j^{-1} x^{2-2r}}{1-x^2}. & \textrm{Magnetic}\\
\end{array} \right.
\end{eqnarray}
By the projection, the chiral letter index changes into
\begin{equation}
f_{chiral}(e^{ia},y_{i},x) = \left\{ \begin{array}{ll}
\displaystyle \sum^{N_f}_{j=1} \frac{y_j^{1} x^{r}     - y^{-1}_j x^{2-r}}{1-x^2}
\left[ \sum^{N}_{i=1}  x^{|m_i|} 2\cos a_i +(-1)^n \right], &\textrm{Electric}\\
\displaystyle \sum^{N_f}_{j=1} \frac{y_j^{-1} x^{1-r}  - y^{ 1}_j x^{1+r}}{1-x^2}
\left[ \sum^{N}_{i=1}  x^{|m_i|} 2\cos a_i +(-1)^n \right]\\
\displaystyle ~~ + \sum^{N_f}_{i=1}  \sum^{N_f}_{j=i} \frac{y_i^{1} y_j^{1} x^{2r}
- y_i^{-1} y_j^{-1} x^{2-2r}}{1-x^2}.&\textrm{Magnetic}\\
\end{array} \right.
\end{equation}

We checked every case discussed in the main text, turning on the chemical potentials, and give one example:
$(N_f,k,N_c) = (2,1,2)$, the electric $O(2)$ and the magnetic $O(3)$,
\begin{eqnarray}
&I&(x,y_1,y_2)\nonumber\\
&=&1+x^2\left(-2-\frac{y_1}{y_2}-\frac{y_2}{y_1}\right)+x^4\left(-1+\frac{y_1}{y_2}+\frac{y_2}{y_1}\right)+\frac{x^{4-2 r}}{y_1 y_2}+x^{3-r}\left(\frac{1}{y_1}+\frac{1}{y_2}\right)\nonumber\\
&&+x^{2 r}\left(y_1^2+y_1 y_2+y_2^2+x^2\left(-2 y_1^2-\frac{y_1^3}{y_2}-2 y_1 y_2-2 y_2^2-\frac{y_2^3}{y_1}\right)\right)\nonumber\\
&&+x^{4 r}\left(y_1^4+y_1^3 y_2+2 y_1^2 y_2^2+y_1 y_2^3+y_2^4\right)+\cdots\nonumber\\
&=& 1-x^2\big(\chi_1(u)+1\big)+x^4\big(\chi_1(u)-2\big)+x^{4-2r}y_0^{-2}+x^{3-r}y_0^{-1}\chi_\frac{1}{2}(u)\nonumber \\
&&+x^{2r}y_0^{2}\left(\chi_1(u)-x^2(\chi_2(u)+\chi_1(u))\right)+x^{4r}y_0^{4}\big(\chi_2(u)+1\big)+\cdots \nonumber\\
\end{eqnarray}
where $y_0=(y_1y_2)^{1/2}$, $u=\frac{y_1}{y_2}$ correspond to the
chemical potentials for global symmetries $U(1)_A \times SU(2)$
respectively.

\end{document}